\documentclass[draft]{agujournal}
\journalname{Geophysical Research Letters}

\begin{document}

\title{Plasma pressures in the heliosheath from Cassini ENA and Voyager 2 measurements: 
       Validation by the Voyager 2 heliopause crossing}

\authors{K. Dialynas\affil{1}, S. M. Krimigis\affil{1,2}, R. B. Decker\affil{2} 
and D. G. Mitchell\affil{2}}

\affiliation{1}{Office of Space Research and Technology, Academy of Athens, Athens 10679, Greece.}
\affiliation{2}{Applied Physics Laboratory, The Johns Hopkins University, Laurel, Maryland 20723, USA.}

\correspondingauthor{Konstantinos Dialynas}{kdialynas@phys.uoa.gr}

\begin{keypoints}
\item ``Ground truth'' Energetic Neutral Atom and ion data in the heliosphere estimate the reported Heliopause crossing by V2 at $\sim$119 AU
\item The normalization of Energetic Neutral Atom and ion intensities yields an interstellar neutral hydrogen density of n$_H\sim$0.12 cm$^{-3}$.
\item The 5.2-24 keV H$^+$ pressures dominate the 5.2-3500 keV distribution whereas pressure balance implies that B$_{ISMF} \sim$ 0.5 nT.
\end{keypoints}

\begin{abstract}
We report ``ground truth'', 28-3500 keV in-situ ion and 5.2-55 keV remotely sensed ENA 
measurements from Voyager 2/Low Energy Charged Particle (LECP) detector and Cassini/Ion 
and Neutral Camera (INCA), respectively, that assess the components of the ion pressure in 
the heliosheath. In this process, we predict an interstellar neutral hydrogen density of 
$\sim$0.12 cm$^{-3}$ and an interstellar magnetic field strength of $\sim$0.5 nT upstream of 
the heliopause in the direction of V2, {\it i.e.} consistent with the measured magnetic field 
and neutral density measurements at Voyager 1 from August 2012, when the spacecraft 
entered interstellar space, to date. Further, this analysis results in an estimated 
heliopause crossing by V2 of $\sim$119 AU, as observed, suggesting that the parameters 
deduced from the pressure analysis are valid. 
The shape of the $>$5.2 keV ion energy spectra play a critical role towards determining the
pressure balance and acceleration mechanisms inside the heliosheath.
\end{abstract}

\section*{Plain Language Summary}
The Voyager missions, together with Cassini, provide the only combination 
of spacecraft to date that can establish ``ground truth'' at $\sim$100 AU and beyond, and 
has recently settled the long standing issue on the dual heliosphere models, 
showing that the heliosphere behaves as rough diamagnetic-bubble.
Leveraging from the synergy between remote sensed ENAs and in-situ measured ions
we estimate (accurately) the recently reported V2
heliopause crossing at $\sim$119 AU, showing that the shape of the ion energy spectra 
play a critical role towards determining the pressure balance and acceleration mechanisms 
inside the heliosheath. In anticipation of measurements from V2 after it crossed the heliopause, 
the normalization of ENA and ion 
intensities provides an important insight on the properties of the Local Interstellar Medium,  
showing an interstellar neutral hydrogen density of n$_H\sim$0.12 cm$^{-3}$ and a magnetic field
upstream of the heliopause of B$_{ISMF} \sim$ 0.5 nT.

\section{Introduction} \label{sec:intro}
For more than half a century, the shape and interactions of the Sun's astrosphere 
(the heliosphere) with the Local Interstellar Medium (LISM) over the solar cycle, have 
been modeled with increasingly sophisticated techniques 
\citep{davis1955,dessler1967,baranov1971,fahr2000,zank2003,opher2004,washimi2007,
izmoenov2008,izmoenov2009,pogorelov2013}. However, none of the past 
theories and models were corroborated by measurements, an inherent limitation that 
was removed only after the first space probes, Voyager-1 and Voyager-2 (V1 \& V2) 
reached the inner boundary of the heliosphere (termination shock, TS) in 2004 and 2007, 
where the supersonic solar wind (SW) terminates at the shock front, at distances of $\sim$94 
\citep{decker2005, stone2005} and $\sim$84 Astronomical Units (1 AU equals the 
distance between Earth and Sun, $\sim$150 million km) \citep{decker2008}, respectively, 
discovering the reservoir of ions and electrons that constitute the heliosheath (HS), 
between the TS and the heliopause (HP).

The two Voyagers are traversing the heliosphere in the upstream (nose) hemisphere, where 
the interstellar flow impinges, and have made two of the key discoveries in heliospheric 
physics during this decade: The heliopause crossings by V1 in 
August of 2012 \citep{krimigis2013, stone2013, burlaga2013} at a distance of $\sim$122 AU, +35$^\circ$ ecliptic latitude 
and the crossing by V2 in November of 2018 (https://www.jpl.nasa.gov/news /news.php?feature=7301) 
at a distance of $\sim$119 AU, -34$^\circ$ from the ecliptic equator.

Remote observations from Cassini (in orbit around around Saturn at
$\sim$10 AU until 15 Sep. 2017) were used to image for the first time the so-called 
``heliotail'' in 2003 \citep{dialynas2015} through its dedicated Energetic Neutral 
Atom (ENA) detector (Ion and Neutral Camera-INCA; \cite{krimigis2004}), providing the first 
full-sky image of the heliosphere in 5.2-55 keV ENAs \citep{krimigis2009} and at $<$6 keV ENAs 
from the Interstellar Boundary Explorer (IBEX) mission, at $\sim$1 AU \citep{mccomas2009}.
In situ measurements of $>$28 keV ions in the heliosheath using the Low Energy Charged Particle (LECP) 
instrument \citep{krimigis1977} on board both Voyagers provided ``ground truth'' to the global ENA 
images through overlapping energy ranges of both ions and neutrals. 

As the ENAs measured by INCA have been shown to originate in the HS 
\citep{dialynas2013, dialynas2017a}, the resulting Cassini/INCA images 
(e.g. Figure \ref{fig:Figure1}a) 
provide a marker for the local plasma-neutral processes inside the heliosheath. 
Figure \ref{fig:Figure1}a shows a ``Belt'' 
of varying ENA intensities, identified as wide ENA region that wraps around the celestial sphere in 
ecliptic coordinates, passing through the ``nose'' the ``anti-nose'' (tail) and the north 
and south heliosphere poles, together with two prominent ``Basins'', identified as two 
extended heliosphere lobes, where the ENA minima occur \citep{krimigis2009,dialynas2013}, 
placing the V1\&2 ion data in a global context. 
The source of the IBEX-defined ``Ribbon'', identified as a bright and narrow stripe of
ENA emissions between the V1 and V2 directions, is thought to lie beyond the 
heliopause \citep{mccomas2017a}, 
with its center coinciding with the direction of the local interstellar magnetic field (ISMF),
but the origin of the IBEX-defined globally distributed flux \citep{livadiotis2011} may
well be the heliosheath \citep{dayeh2011}, as also inferred in \citep{dialynas2013}.

The combination of remotely imaged 5.2-55 keV INCA/ENAs, together with $>$40 keV in-situ 
ion measurements from the Low Energy Charged Particle (LECP) experiment on V1 
\citep{decker2005} in the heliosheath (HS) have been used in the past to 
predict the V1 heliopause crossing \citep{krimigis2011} and the magnitude of 
the interstellar magnetic field \citep{krimigis2010} with good accuracy. Key discoveries 
through the LECP experiment's measurements of $>$28 keV (V2) ions, taken together with the 5.2-55 keV 
INCA/ENAs, showed that the heliosphere responds promptly, within $\sim$2-3  years, to outward 
propagating solar wind changes in both the nose and tail directions over the solar cycle 
and suggested a diamagnetic ``bubble-like'' heliosphere with few substantial tail-like 
features \citep{dialynas2017a,dialynas2017b}. This bubble heliosphere concept is 
consistent with recent advanced modeling 
\citep{opher2015,drake2015,kivelson2013,golikov2017,opher2019} as well as ENA 
observations from the IBEX mission \citep{galli2016,galli2017}, and has settled the issue on the 
dual heliosphere models first posited by \cite{parker1961} over five decades ago, 
concerning the properties and time evolution of the heliosphere and its interaction 
with the Local Interstellar Medium (LISM).

\section{5.2-3500 keV Energy Spectra in the Heliosheath} \label{sec:spectra}
\begin{figure}[ht!]
  \noindent\includegraphics[width=\hsize]{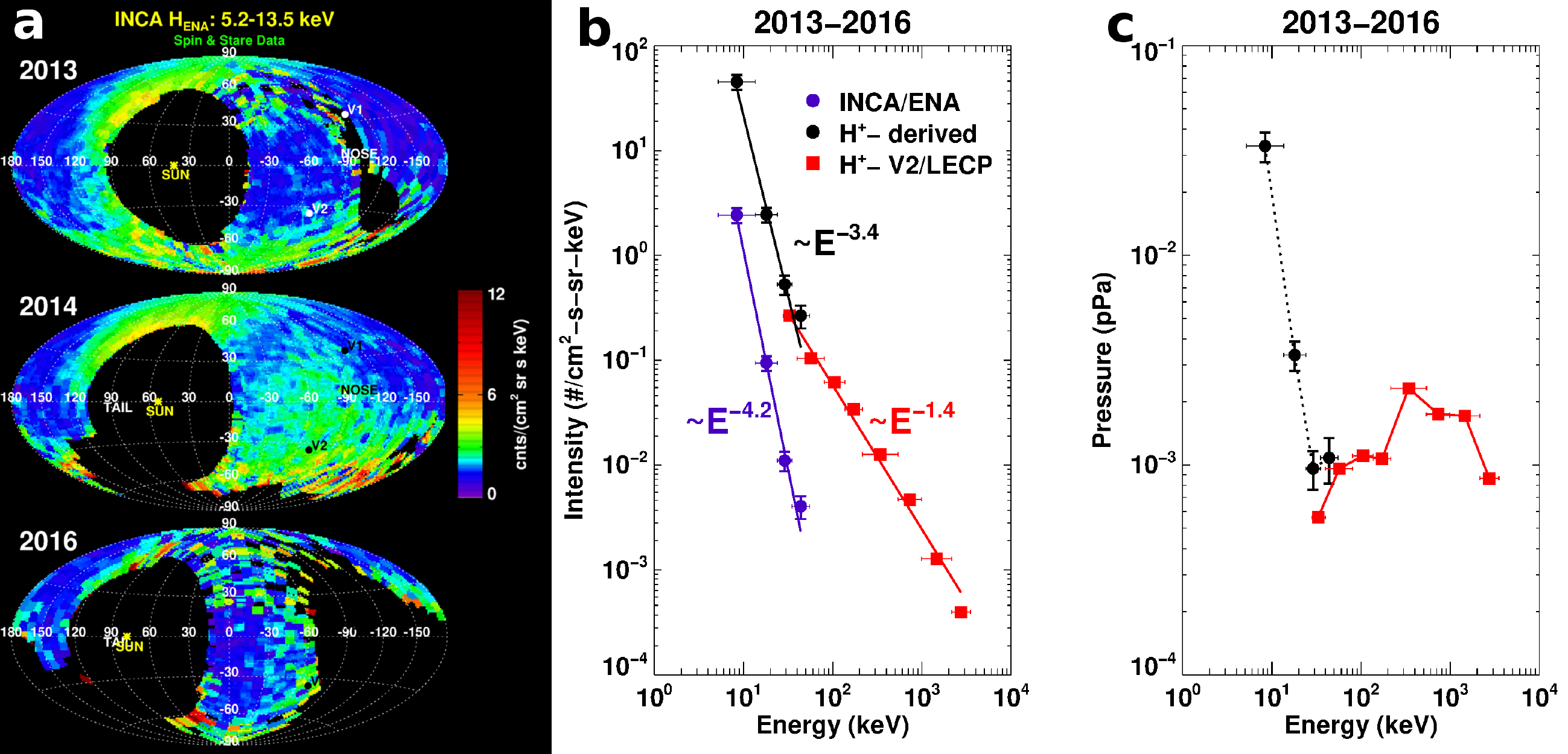}
  \caption{(a) A combination of 320x160 pixel INCA/ENA images (5.2-13.5 keV) organized 
  in ecliptic coordinates, over the 2013-2016 time period, after the ENA and ion minimum, 
  that corresponds to the onset of SC24. (b) Average 5.2-55 keV ENA energy spectra of 
  the INCA/ENA data in the pixels enclosing the position of Voyager 2 (5$^\circ$x5$^\circ$), 
  together 
  with the deduced H$^+$ spectra and the 28-3500 keV ion energy spectra measured in-situ by Voyager 
  2/LECP measurements over the time period 2013-2016. Horizontal bars indicate the 
  INCA and LECP energy passbands for H ENAs and ions, respectively. The spectra 
  are fitted with a power law form in energy in a least square sense; relative percentage 
  errors in the spectral slope, do not exceed 8\%. 
  (c) The 5.2-3500 keV H$^+$ pressure energy spectra inside the heliosheath, derived
  from the measurements shown in panel (b), using L$_{HS}$=35.2 AU and n$_{H} \sim$0.12 cm$^{-3}$.}
  \label{fig:Figure1}
\end{figure}

ENAs are products of charge exchange (CE) \citep{lindsay2005} between 
fast protons and the ``background'' neutral hydrogen (H) gas flowing through the 
heliosheath \citep{krimigis2009,mccomas2009}. Due to overlapping 
energy bands between INCA and LECP we are able to deduce with certainty the nonthermal energetic 
ion contribution in the overall HS dynamics (Figure \ref{fig:Figure1}b,c). 
Thus, the normalization (see \cite{krimigis2009}) of the intensity of the 
highest ENA energy channel ($\sim$35-55 keV) 
measured remotely at $\sim$10 AU to the lowest V2 H$^+$ channel ($\sim$28-40 keV) making in-situ 
ion measurements inside the HS, yields a HS thickness along the V2 trajectory of 
L$_{V2}\sim$(35.2 $\pm$ 8.6) AU (the uncertainty in L$_{V2}$ is calculated 
from the error propagation function due to the measured uncertainties in the ENA and 
ion intensities), assuming a neutral Hydrogen density of n$_H\sim$0.12 cm$^{-3}$, 
and suggests a HP crossing at $\sim$119.2 AU (Figure \ref{fig:Figure1}b). 
In November of 2018 V2 crossed the HP at a distance 
of $\sim$119 AU, indicating that our calculation is not only relevant, but it once again highlights 
that the source of the 5.2-55 keV ENAs detected with Cassini/INCA is the HS.

\cite{roelof2012} showed that consideration of the Compton-Getting factor \citep{compton1935} 
in ENA and ion measurements in V2 (and V1) over the time period before 2012, may increase 
the estimate of the HS radial thickness by some percentage. The radial plasma velocities 
in V2, however, are decreasing, from $\sim$80-90 km/s in 2013 to gradually 
approaching zero values towards 2016. Taking an average radial plasma speed of V$_r \sim$40 km/s 
and a spectral index for the lower LECP channels of $\gamma \sim$1.7 (see Figure \ref{fig:Figure3}a), 
then the estimated heliosheath width in the V2 direction (L$_{V2}$) is related to the 
Compton-Getting corrected value (L$^{'}_{V2}$) as 
L$^{'}_{V2}$/L$_{V2}$ = (1 + V$_r$/V)$^{2(\gamma+1)}$ = (1 + 40/2520)$^{5.4}$ = 1.088 \citep[e.g][]{roelof2012}. 
This number indicates that our estimate of the heliosheath width (L$_{V2}\sim$35.2 AU) 
can be increased by $\sim$8.8\%, which translates
to $\sim$3.1 AU, i.e. much smaller than the calculated error bar ($\pm$ 8.6 AU). 
Therefore, although the Compton-Getting correction has been initially considered, it was 
found to be small because of the low velocities in the HS during the time period in 
question. Overall, these ENAs serve as important indicators of the acceleration processes 
that the parent H$^+$ population undergoes inside the HS, thus imposing a key constraint 
on any future interpretation concerning the HS dynamics.

Although the Plasma (PLS) instrument on V1 \citep{bridge1977} failed in 1980, the Plasma Wave (PWS) instrument 
\citep{scarf1977} is in full operational condition, thus detecting electric field 
emissions, which can be related to the electron density from the frequency of 
electron plasma oscillations. Assuming that the 
equilibrium ionization fraction, n$_e$/(n$_e$+n$_H$), is $\sim$ 50\% for the LISM, then 
the neutral hydrogen (n$_H$) density is directly comparable to the measured electron density n$_e$. 
Consequently, with V1 traversing the LISM since 2012, the neutral 
hydrogen density upstream of the HP has been indirectly measured \citep{gurnett2013, gurnett2015} 
to be $\sim$0.09-0.11 cm$^{-3}$ (although densities up to $\sim$0.14 cm$^{-3}$ were also found at 
distances $\sim$20 AU past the HP as reported in \cite{gurnett2017}), i.e. 
consistent with the 0.12 cm$^{-3}$ that was used here. For clarity, we repeated the 
calculation after assigning a neutral density of $\sim$0.1 cm$^{-3}$, showing a HS thickness of 
L$_{V2} \sim$(42.2 $\pm$ 10.3) AU, which is roughly consistent with the V2 HP crossing within the 
calculated uncertainties. Past observations from Ulysses/SWICS \citep{gloeckler2001} and other 
measurements [see \cite{bzowski2008}] were also consistent with values about 0.1 $cm^{-3}$.

At this point we cannot determine if there is a possible 
density gradient between the V1 and V2 LISM locations along the HP boundary, or if the 
inferred n$_{H} \sim$0.12 cm$^{-3}$ in the V2 direction is only a manifestation of the wide range of 
densities, that were found to be increasing from $\sim$0.09 to 0.14 cm$^{-3}$ radially outward along 
the V1 trajectory, upstream of the HP. In principle, an electron density gradient does 
not necessarily imply a corresponding neutral Hydrogen density gradient, which is affected 
by the addition of neutrals via upstream charge exchange between the deflected plasma 
protons flowing around the HP and the incoming interstellar neutral H-atoms, thus forming 
a rather weak ``hydrogen wall'', which then depends on the presence and strength of a bow 
shock upstream of the HP (see \cite{dialynas2017a} and references therein).

The shape of the ion energy spectra play a critical role towards determining the pressure 
balance and acceleration mechanisms inside the heliosheath. The average ENA energy 
spectra in Figure \ref{fig:Figure1}b are consistent with a power-law form in energy 
(J$_{ENA} \sim$E$^{-(4.2 \pm 0.2)}$), 
whereas the resulting ENA-derived H$^+$ spectrum is less steep 
(J$_{ENA-derived H+} \sim$E$^{-(3.4 \pm 0.2)}$) 
because of the energy dependence of the CE cross sections, as explained in 
\citep{krimigis2009}. Recent observations from the New Horizon spacecraft at $\sim$38 AU 
\citep{mccomas2017b} showed that the pick-up ion distribution is heated in the 
frame of the solar wind with increasing distance, before reaching the TS region 
at $\sim$90 AU. Although the TS was considered to be a site at which Anomalous Cosmic Rays (ACRs) 
are accelerated, the $\sim$10-100 MeV intensities in both V1 \& V2 did not peak at the TS as 
expected \citep{stone2005,stone2008}. Contrary to expectations, the shocked thermal 
plasma upstream of the TS remained supersonic, as only 20\% 
of the upstream energy density went into heating the downstream thermal plasma 
\citep{richardson2008}. The rest of the SW energy was transferred into heating pickup 
ions (PUI) and $>$15\% transferred to the $>$28 keV protons. This is translated to a prominent
hardening break (less steep spectrum) in the $>$28 keV part of the H$^+$ distribution (e.g. Figure \ref{fig:Figure1}b) 
that was attributed to an accelerated ``core'' interstellar pickup ion distribution at the 
TS, through shock drift acceleration and particle scattering in the vicinity of the shock 
\citep{giacallone2010}, as one of the possible mechanisms.

This characteristic seems to persist throughout the heliosheath as shown in Figure \ref{fig:Figure1}b, where 
the $>$28 keV spectra fit smoothly to the ENA-derived H$^+$ spectra at the energy range of 
$\sim$24-80 keV, but the overall 28-3500 keV ion spectra exhibit a rough power law form in 
energy with J$_{LECP}\sim$E$^{-(1.4 \pm 0.1)}$. As explained in \cite{dialynas2013}, 
the INCA spectra
exhibit hardening breaks at $>$35 keV (e.g. Figure \ref{fig:Figure1}b), which, due to the uncertainties 
related to the INCA/ENA measurements, are accounted as not statistically significant 
(therefore, the spectra can be described by a single power-law function that applies 
to the whole INCA energy range, as was also shown in Figure \ref{fig:Figure1}b). However, a simple power 
law fit in the 24-55 keV ENA-derived H$^+$ intensities, shows that the spectra are consistent with a 
$\sim$E$^{-(1.7 \pm 0.8)}$ law in this energy range. At the same time, the V2/LECP ion spectra 
exhibit a turn-up in the intensities at the energy range of 28-80 keV, thus informing 
that the change in the power law slope over the whole 5.2-3500 keV distribution 
(hardening break) occurs within the 24-80 keV energy range. Interestingly, the 28-80 keV 
LECP distribution follow a $\sim$E$^{-(1.7 \pm 0.1)}$ law, i.e. both the 24-55 keV ENA-derived H$^+$ 
measurements and the 28-80 keV V2/LECP ones have the same rough slope (Figure \ref{fig:Figure3}a).

Despite the $>$140 AU separation between the two Voyagers (+35$^\circ$ and −34$^\circ$ latitude, 
respectively) since they both entered the heliosheath to date (they are $\sim$165 AU apart today), 
the ion spectra at V1 and V2 inside the heliosheath are very similar in both 
shape and number as a function of time (e.g. \cite{decker2009} and also Figure \ref{fig:Figure2}c,d, 
this study). In addition, they are in good agreement with the INCA/ENA data when converted to H$^+$ 
using standard parameters explained earlier, in overlapping energy bands 
(e.g.\cite{dialynas2017a} and Figure \ref{fig:Figure2}c,d this study).

\section{Pressure Balance in the Heliosheath} \label{sec:pressure}
After the V1 and V2 respective crossings of the TS, it was found that the heliosheath 
pressure is dominated by suprathermal particles. While the $>$28 keV partial pressure 
distribution is measured in-situ by LECP, we use the ENA measurements converted to ions in the 
HS to compute the partial plasma pressure at $>$5.2 keV 
($P(dynes/cm^{2}) = (8\pi/3)(m/2)^{1/2} J_{ion} E^{1/2} \Delta E$, where $E=\sqrt{E_{1} \cdot E_{2}}$ 
is the midpoint of the measured energy in each energy channel, $E_{1}, E_{2}$ are each 
channel passbands, $\Delta E = E2-E1$, m is the proton mass and $J_{ion}$ is the proton intensity; note 
that by substituting $p=mv$ in this equation, we obtain 
$\Delta P= (4πp/3) J_{ion} \Delta E$, as used in \cite{dialynas2015}), a range where 
many of the PUIs associated with the TS and heliosheath reside. The 5.2-24 keV H$^+$ pressures 
shown in Figure \ref{fig:Figure1}c dominate the 5.2-3500 keV pressure distribution, which indicates that the 
5.2-55 keV part of the energetic H$^+$ distribution covered by the Cassini/INCA is critically
important for determining the pressure balance inside the heliosheath and cannot be neglected.

\begin{figure}[ht!]
  \noindent\includegraphics[width=\hsize]{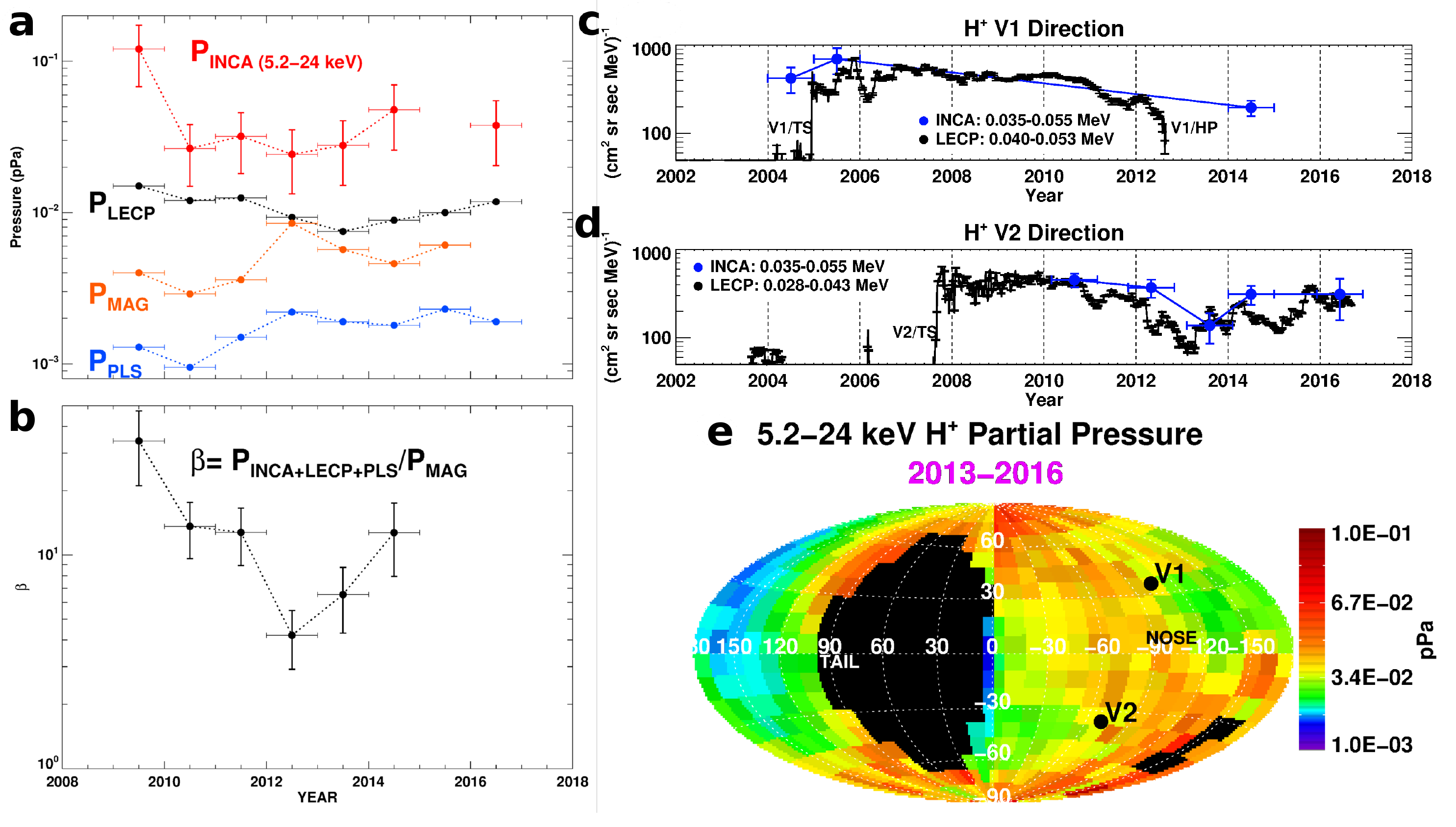}
  \caption{(a) Yearly averaged pressure profiles of (red line) remotely sensed 
  5.2-24 keV ENA-derived H$^+$ over 5$^\circ$x5$^\circ$ enclosing the V2 pixel, (black line) 
  28-3500 keV H$^+$, (blue line) $>$10 eV H$^+$ and (orange line) 
  magnetic field from Cassini/INCA, V2/LECP,
  V2/PLS and V2/MAG experiments, respectively, as a function of time from 2009 to 2016.
  (b) Yearly H$^+$ partial pressure (PLS, INCA, LECP) divided by the magnetic field pressure (MAG) 
  inside the heliosheath for the 2009-2014 time period. (c) 35-55 keV INCA/ENA measurements 
  averaged over 5$^\circ$x5$^\circ$ enclosing the V1 pixel and converted to ion 
  intensities using L$_{V1}\sim$28 
  AU and n$_{H}\sim$0.1 cm$^{-3}$, compared directly with the in-situ 40-53 keV LECP ion histories 
  (see \cite{dialynas2017a} for details). (d) The same as (c) for the $>$24 keV INCA/ENA 
  measurements around the V2 pixel, using L$_{V2}\sim$35.2 AU and n$_{H}\sim$0.12 cm$^{-3}$, 
  as derived in 
  Figure \ref{fig:Figure1}b, compared directly with the 28-43 keV LECP ion histories at V2. 
  (e) 5.2-24 keV pressure contributed by H$^+$ inside the heliosheath 
  (between the TS and the HP) computed 
  from spectra deduced from the ENA observations using a varying heliosheath thickness 
  and hydrogen density towards the upstream (nose) hemisphere, as detailed in the text.
  The mean relative percentage error is $\sim15\%$}
  \label{fig:Figure2}
\end{figure}

The H$^+$ partial pressure from the 5.2-24 keV INCA channel is a factor of $\sim$4 higher 
than the $>$28 keV LECP pressure (on average) throughout the 2009-2016 time period and a 
factor of $\sim$30 higher than the measured PLS thermal pressure over the same time period 
(Figure \ref{fig:Figure2}a). The partial plasma beta shown in Figure \ref{fig:Figure2}b ($\beta = P_{particle}/P_{MAG}$) 
inside the HS is 
persistently $>>$4 on average (a local minimum that corresponds to the minimum of SC23 with 
a time delay of $\sim$2-3 years as explained in \cite{dialynas2017a,dialynas2017b}) pointing towards a 
heliosphere that exhibits diamagnetic behavior. Although magnetic field measurements for the 
year 2016 are not yet available, if we assign a magnetic field strength at V2 in 2016 
of the order of B$\sim$0.14 nT (roughly comparable to 2015), we obtain 
$\beta_{2016}\sim$6.6 over 2016. Here we should note 
that the calculations in Figure \ref{fig:Figure2}b do not take into account the partial 
pressure that corresponds to the IBEX energy range, that would result in even higher 
numbers for the plasma-$\beta$, and would further support the arguments provided in this study.

The belt is a relatively stable feature as a function of energy, and corresponds to the 
reservoir of particles inside the HS, constantly replenished by new particles from the SW. 
As also argued in \cite{krimigis2010} and \cite{dialynas2015}, Figure \ref{fig:Figure2}e 
demonstrates that the belt ENAs are associated with a region of enhanced particle pressure that is 
formed between the TS and the HP and contribute significantly to balancing the pressure of 
the ISMF. Although only part of the anti-nose (downstream hemisphere) is covered in the 
2013-2016 INCA/ENA measurements, the partial 5.2-24 keV H$^+$ pressure is roughly comparable
between the upstream and downstream hemispheres, at least in the regions where the 
belt is identified, whereas the partial pressure at the basins are a factor of $\sim$6-7 lower.

The overall pressure distribution in Figure \ref{fig:Figure2}e, taken together with the time variant pressure 
distributions shown in Figure \ref{fig:Figure2}a, the ENA and ion intensities shown in 
Figure \ref{fig:Figure2}c and the 
$\beta$-parameter shown in Figure \ref{fig:Figure2}b are consistent with the concept of a roughly symmetric 
HS that behaves as a diamagnetic bubble, as shown in the conceptual model of \cite{dialynas2017a}. 
Although, as noted earlier, we cannot determine 
with certainty the possibility of a neutral hydrogen density gradient between the V1 and V2 LISM locations 
along the HP boundary, the pressures shown in Figure \ref{fig:Figure2}e are computed from spectra deduced 
from the ENA observations using a varying heliosheath thickness and hydrogen density towards 
the upstream (nose) hemisphere: $\sim$35 AU and 0.12 cm$^{-3}$ over the -90$^\circ$ to 
-30$^\circ$ in latitude (consistent with the V2 HP crossing), $\sim$28 AU and 0.1 cm$^{-3}$ 
over +30$^\circ$ to +90$^\circ$ in latitude (consistent with the V1 measured parameters) 
and  $\sim$31 AU and 0.11 cm$^{-3}$ over -30$^\circ$ to +30$^\circ$ in 
latitude (to compensate for a possible density gradient).
Despite these uncertainties, the overall 5.2-24 keV partial pressure around the V1 and V2 pixels 
(Figure \ref{fig:Figure2}e) is $\sim$0.033 pPa, whereas the peak to basin partial pressure 
(belt to basins, respectively) in Figure \ref{fig:Figure2}e is within the range of 
$\sim$0.092-0.014 pPa.

The measurements shown here can be used to address the pressure balance at the 
interaction region between the HS and the LISM, i.e. the heliopause. On average, the 
partial 0.7-4.3 keV H$^+$ pressure in the V2 (and V1) direction from IBEX is found to be 
$\sim$27 pdyn AU cm$^{-2}$ \citep{mccomas2014} and assuming a HS thickness of $\sim$35 AU this 
yields P$_{0.7-4.3keV}\sim$0.077 pPa. At higher energies, the 5.2-24 keV partial pressure fluctuates 
about $\sim$0.025 to 0.105 pPa over the 2009-2016 time period, with an average value of 
P$_{5.2-24keV}\sim$ 0.05 pPa, whereas the V2/LECP partial pressure is P$_{>28 keV}\sim$0.013 pPa 
(ranging from 0.008 to 0.016 pPa over 2009-2016). The magnetic field pressure is 
much smaller, i.e. P$_{MAG}\sim$0.005 pPa whereas the thermal pressure is 
also $\sim$0.005 pPa \citep{krimigis2010}. 
Thus, the overall (isotropic) pressure in the heliosheath is calculated by adding the 
aforementioned partial pressures, i.e. P$_{HS}\sim$ 0.1522 pPa. 
\cite{krimigis2010}, using measurements from V1 and V2 immediately 
downstream of the TS, presented the reasonable assumption that despite 
possible adiabatic cooling throughout the HS, this 
pressure would be carried out to the HP and that the thermal ram pressure will not affect 
the force balance at the HP (as there should be no flow across an ideal heliopause). Here we 
use average pressure measurements from inside the HS (from 2009) towards the HP (up to 2016)
around the V2 pixel.

Neglecting the magnetic tension stress, and assigning P$_{IS}$(thermal) =0.01 pPa and 
P$_{IS}$(dynamic) = 0.0565 pPa (adopted from \cite{krimigis2010}), then ρV$^2$/2 + P + B$^2$/2$\mu_0$ should 
be constant along the flow streamline ($\mu_0$ = 4$\pi$x10$^{−7}$ H/m, magnetic permeability), which 
means that the IS magnetic field pressure is 
P$_{ISMF}$ $\sim$ P$_{HS}$ - [P$_{IS}$(thermal) + P$_{IS}$(dynamic)] = 0.0857 pPa, thus, 
providing an estimate of 
the IS magnetic field strength to be B$_{ISMF}$ $\sim$0.47nT. This number is the result of a rough 
estimate of the pressures inside the heliosheath and subject to parameters 
that are not accurately known in the upstream medium, but is consistent with the predicted 
magnetic field upstream of the HP that is derived from recent sophisticated modeling 
\citep{opher2019}. Further, previous estimates of 
the IS magnetic field using the 2003-2009 INCA measurements predicted B$_{ISMF}<$0.6 nT along the 
V1 direction \citep{krimigis2010} that were confirmed \citep{burlaga2016} after 
the V1 crossing of the HP. Although the magnetic field measurements upstream of 
the HP from V2 have not yet become available, and might differ from our 
estimate, these numbers are very close to the V1 measurements where B$_{ISMF}$ fluctuated 
about $\sim$(0.48 $\pm$0.04) nT from 2012 \citep{burlaga2016} to date, i.e. up to at least 25 AU 
past the HP.

\section{Discussion} \label{sec:discussion}
We have demonstrated that the 5.2-55 keV INCA/ENA measurements, originating in the 
heliosheath, can be used to estimate (accurately) the recently reported V2 heliopause crossing 
at $\sim$119 AU and delineate the components of the ion pressure in the heliosheath. We have 
also argued that those measurements are critically important for determining the 
pressure balance in the heliosheath, providing realistic numbers for the interstellar 
neutral Hydrogen density and magnetic field. Following the arguments provided in this 
study, we can further explore the consequences of underestimating and/or neglecting 
the shape/intensities of the 5.2-55 keV spectra from Cassini/INCA.

\begin{figure}[ht!]
  \noindent\includegraphics[width=\hsize]{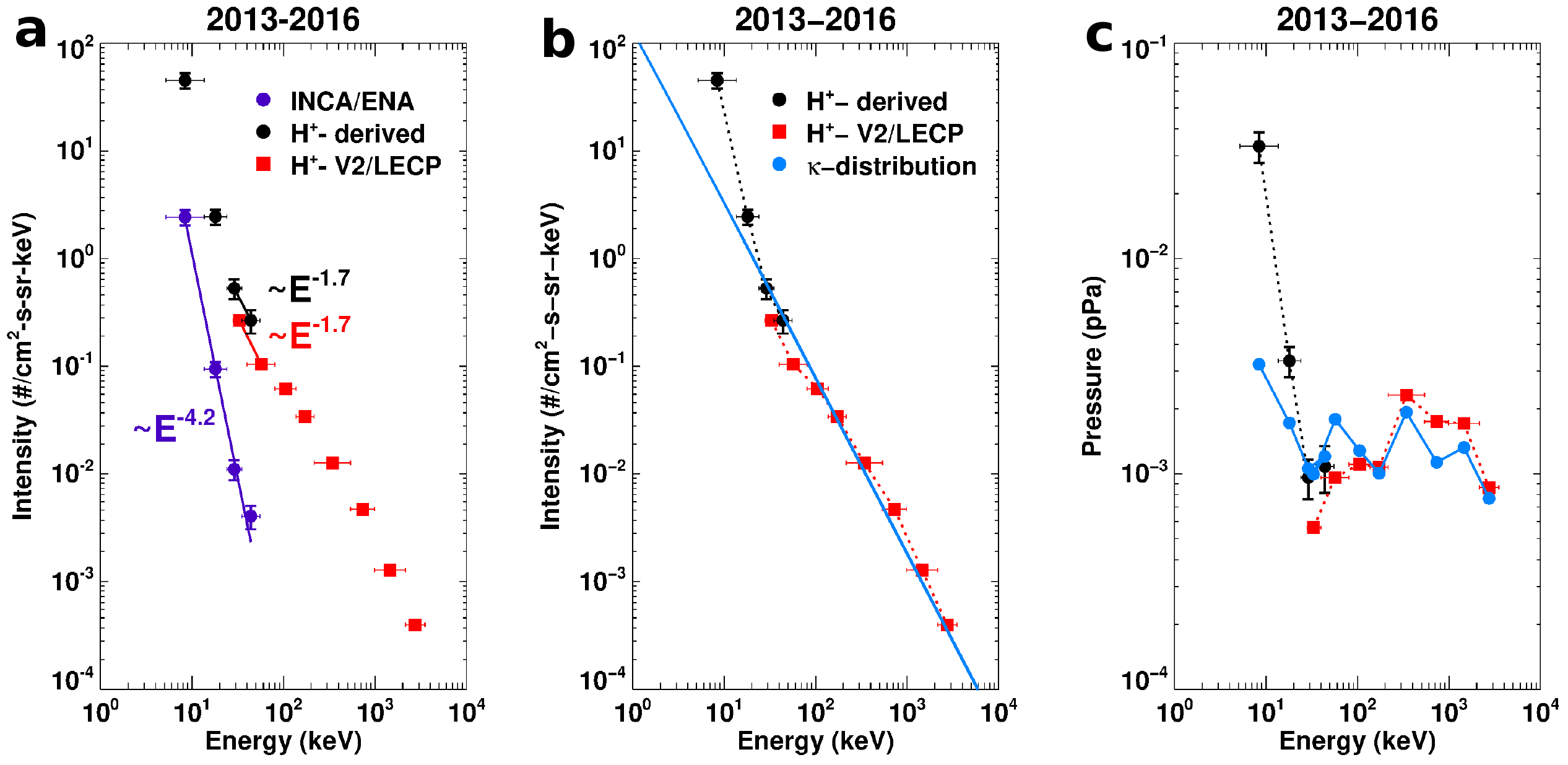}
  \caption{(a) The same measurements shown in Figure \ref{fig:Figure1}b fitting (black and red lines)
  only the 24-55 keV 
  INCA-derived H$^+$ and the 28-80 keV V2/LECP ion data (black and red points, respectively), to demonstrate that these 
  populations retain the same spectral slopes ($\sim$E$^{-1.7}$), as detailed in the text. (b) Same 
  measurements as in (a), incorporating a single $\kappa$-distribution that spans over the 
  eV to MeV energy range with  T$_p$=0.26 keV (=3 x 10$^6$ K), n$_p$=0.002 cm$^{-3}$ and 
  $\kappa$=1.63 \citep{zirnstein2015,zirnstein2018} as detailed in the text. 
  (c) The same as in Figure \ref{fig:Figure1}c, incorporating the pressure that results from the 
  $\kappa$-distribution shown in panel (b), calculated in the positions of the INCA and LECP channels.}
  \label{fig:Figure3}
\end{figure}

Although $\kappa$-distributions are very useful towards characterizing the ion spectra in space 
plasmas \citep{dialynas2017c}, the overall shape of the $>$5.2 keV spectra deviates 
substantially from any simplified notion that may include a single $\kappa$-distribution 
to describe the particle spectra from eV to MeV energies, even 
if selected as an initial condition at the TS site that will, subsequently, be subjected 
to charge-exchange and velocity diffusion inside the heliosheath and may eventually 
roughly resemble the spectra shown in Figure \ref{fig:Figure1}c. For example, the V2/LECP 
ion spectra may be consistent with a $\kappa$-distribution (e.g. \cite{zirnstein2015,zirnstein2018}) 
using T$_p$=0.26 keV (=3 x 10$^6$ K) \citep{heerikhuisen2008}, 
n$_p$=0.002 cm$^{-3}$ \citep{richardson2014} and $\kappa$=1.63 \citep{decker2005} at the TS and 
inside the HS. Although such an approach would likely fit the multi-hundred keV high energy tails 
measured by LECP with good accuracy, thus providing an adequate representation of 
this partial pressure, it would, at the same time, undershoot the 5.2-24 keV part of 
the H$^+$ distribution (Figure \ref{fig:Figure3}b,c). Specifically, the modelled H$^+$ pressure over the 2013-2016 
time period in the 5.2-13.5 keV INCA channel would become $\sim$0.00322 pPa (whereas the measured 
5.2-13.5 keV H$^+$ pressure is $\sim$0.033 pPa, i.e. a factor of $\sim$10.3 higher). In the same manner, 
the modelled H$^+$ pressure in the 13.5-24 keV INCA channel would become $\sim$0.0017 pPa 
(whereas the measured 13.5-24 keV H$^+$ pressure is $\sim$0.0033 pPa, i.e. a factor of $\sim$1.9 higher).

Evidently, by assuming a $\kappa$-distribution, the overall 5.2-24 keV 
pressure will be underestimated by a 
factor of $\sim$6, and the 2009-2016 partial INCA pressure would become P$_{5.2-24 keV}\sim$0.0083 pPa.
Then the P$_{HS}$ is $\sim$0.109 pPa and P$_{IS}\sim$0.042 pPa, which in turn would give 
B$_{ISMF}\sim$0.33 nT, i.e. at least a factor of 1.6 lower than the measured magnetic field from V1, 
$\sim$0.48 nT (and a factor of 1.9 lower than the magnetic field measured by V1 
immediately upstream of the HP, $\sim$0.6 nT, inside the ``pile-up'' region). Further, if one 
completely neglects the contribution of the 5.2-24 keV partial pressure to the overall pressure 
distribution inside the HS, then B$_{ISMF}\sim$0.29 nT. In the same manner, the $\beta$-parameter results much 
lower than unity, if only the PLS measurements are included.

Clearly, $\sim$40\% of the 0.7-24 keV partial pressure ($\sim$0.127 pPa) in the V2 direction is 
accounted for by the 5.2-24 keV part of the ion distribution ($\sim$0.05 pPa). Underestimating 
the partial particle pressure inside the HS, either due to a simplified model 
for the spectral shape that underestimates the 5.2-24 keV ion intensities, or neglecting 
the pressure that comes from this part of the distribution for whatever reason, 
results in B$_{ISMF}$ values of $\sim$0.29-0.33 nT that are frequently used in 
heliosphere models as an upper limit (e.g. \cite{bzowski2017}). The combination of these 
values for the magnetic field together with substantially lower neutral densities upstream 
of the HP (e.g. 0.067 cm$^{-3}$) to characterize the region immediately outside the HP, point 
to comet-type tails concerning the shape of the global heliosphere. These comet-type tails are
contrary to observations that stem from both INCA (a rough “bubble”; \cite{krimigis2009,dialynas2017a}) 
and IBEX (either a rough “bubble” as in \cite{galli2016,galli2017} or an 
“intermediate situation” as in \cite{mccomas2013}), and with recent 
magnetohydrodynamic models \citep{opher2015,drake2015,kivelson2013,izmoenov2015,opher2019} 
concerning the heliospheric configuration.

\acknowledgments
This work was supported at JHU/APL by NASA under contracts NAS5 97271, NNX07AJ69G and NNN06AA01C and 
by subcontract at the Office for Space Research and Technology. The authors are 
grateful to all Cassini/MIMI and Voyager/LECP team members for useful discussions that
made this work possible, and to E. C. Roelof for incisive comments. 
The authors are particularly grateful to M. Kusterer 
for software development and assistance with the Cassini/INCA data processing.
The Cassini/MIMI and Voyager 1 \& 2 LECP measurements, including the INCA/ENA and in-situ 
LECP ion data used in this study can be accessed through NASA public Planetary Data System 
(PDS: https://pds.nasa.gov/) together with the corresponding user guides. 


\end{document}